\documentclass[a4paper,11pt]{article}
\usepackage[inline]{enumitem}
\usepackage[multidot]{grffile}
\usepackage{graphicx}
\usepackage{subfigure}
\usepackage[margin=1in]{geometry}
\usepackage{dcolumn}
\usepackage{bm}
\usepackage{amsmath}
\usepackage{amsthm}
\usepackage{framed}
\usepackage{soul}
\usepackage{amsfonts}
\usepackage{bbold}
\usepackage{dsfont}
\usepackage{amssymb}
\usepackage{color}
\usepackage{latexsym}
\usepackage{stackrel}
\usepackage{slashed}
\usepackage{empheq}
\usepackage{fancybox}
\usepackage{pstricks}
\usepackage{indentfirst}
\usepackage{mathrsfs}
\usepackage{tablefootnote}
\usepackage{longtable}
\usepackage{multirow}
\usepackage{epsfig,psfrag}
\usepackage{subfigure}
\usepackage{mathtools}
\usepackage{setspace}
\usepackage[utf8]{inputenc}
\usepackage[scientific-notation=true]{siunitx}
\usepackage{verbatim}
\usepackage{comment}
\usepackage[many]{tcolorbox}
\usepackage{JHEPpub}
\usepackage{JHEPorcid}
\usepackage[T1]{fontenc}
\newcommand{\bk}{{\boldsymbol k}}

\newcounter{subfiggroup}
\newcommand{\savedthesubfigure}{}

\graphicspath{fig/}
\title{$\Delta$-resonance contributions to QED radiative corrections in neutron and inverse beta decay}
\author[a,b]{Oleksandr Tomalak\orcidlink{0000-0002-4827-5842}}
\author[a,b,c,d]{Yi-Bo Yang\orcidlink{0000-0002-3213-0438}}
\emailAdd{tomalak@itp.ac.cn}
\affiliation[a]{Institute of Theoretical Physics, Chinese Academy of Sciences, Beijing 100190, China}
\affiliation[b]{University of Chinese Academy of Sciences, School of Physical Sciences, Beijing 100049, China}
\affiliation[c]{School of Fundamental Physics and Mathematical Sciences, Hangzhou Institute for Advanced Study, UCAS, Hangzhou 310024, China}
\affiliation[d]{International Centre for Theoretical Physics Asia-Pacific, Beijing/Hangzhou 100190, China}
\abstract{We incorporate the $\Delta(1232)$ resonance into pion-induced QED radiative corrections to neutron decay and inverse beta decay (IBD). Within the framework of heavy-baryon chiral perturbation theory with explicit $\Delta$ degrees of freedom, we compute additional contributions and study their impact on IBD cross sections and on the renormalization of the nucleon isovector vector and axial-vector charges. $\Delta$ resonance does not renormalize the vector charge. For the axial-vector charge, including the $\Delta$ resonance improves convergence and reduces the QED radiative correction to the experiment-over-lattice-QCD ratio $g_A/\left(g^\mathrm{QCD}_A g_V\right)$. $\Delta$ resonance increases the pion-induced QED radiative corrections to IBD by a factor $1.2$-$1.3$.}


\begin{document}
\maketitle
%

\section{Introduction}

The inverse beta decay (IBD) process \(\bar{\nu}_e + p \rightarrow e^+ + n\) is the primary detection mechanism for reactor and supernova antineutrinos in water-based and hydrocarbon-based detectors. With next-generation experiments such as JUNO~\cite{JUNO:2015zny,JUNO:2015sjr,JUNO:2021vlw}, Hyper-K~\cite{Hyper-KamiokandeProto-:2015xww,Hyper-Kamiokande:2018ofw}, and DUNE~\cite{DUNE:2020ypp,DUNE:2021tad} targeting percent- and sub-percent-level precision in neutrino oscillation parameters, an accurate theoretical description of the IBD cross section is a necessary input. Achieving precision demands a systematic treatment of electroweak, strong-interaction, and quantum electrodynamics (QED) radiative corrections within a controlled theoretical framework.

In the preceding papers~\cite{Tomalak:2025jtn,Tomalak:2025okl}, we performed a detailed evaluation of QED radiative corrections to IBD within the framework of heavy-baryon chiral perturbation theory (HBChPT). Subsequently, we included explicit pion degrees of freedom in~\cite{Tomalak:2026btz}. The analysis demonstrated that pion-induced electromagnetic corrections contribute at the sub-percent level to the IBD cross section, with the dominant kinematic dependence arising from leading-order pion loops in the HBChPT expansion. At next-to-leading order, recoil effects and the HBChPT Wilson coefficient \(c_4\) produce only moderate modifications, well below the uncertainties from the nucleon form factors for (anti)neutrino energies up to several hundred MeV.

\(\Delta(1232)\) resonance plays a crucial role in (anti)neutrino-nucleon interactions at low and intermediate energies. In processes such as neutron decay and IBD, \(\Delta\) can make significant contributions through pion-exchange and radiative-correction diagrams that are not captured by the effective theory with pions only. The chiral perturbation theory with explicit \(\Delta\) degrees of freedom has been extensively developed over the past decades~\cite{Jenkins:1990jv,Hemmert:1996xg,Hemmert:1996rw,Hemmert:1997ye,Kambor:1997ns,Fettes:2000bb,Pascalutsa:2002pi,Geng:2008mf,Krebs:2009bf}. This framework provides a systematic power counting for observables involving the \(\Delta\) resonance, treating the \(\Delta\)-nucleon mass splitting \(\Delta_0 = m_\Delta - m_N \sim 293~\text{MeV}\) as a small parameter compared to the chiral symmetry breaking scale. The inclusion of \(\Delta\) has proven essential for achieving convergence in calculations of the nucleon isovector axial-vector charge \(g_A\)~\cite{Procura:2006gq,Hall:2025ytt,Alvarado:2021ibw,Hall:2025ytt,Alvarado:2026kny}, the pion-nucleon scattering amplitudes~\cite{Hoferichter:2015tha,Siemens:2016jwj,Siemens:2017opr,Siemens:2020vop}, and various electromagnetic observables~\cite{Scholten:2002tn,Hemmert:1996xg,Hemmert:1996rw,Pascalutsa:2002pi,Pascalutsa:2005vq,Pascalutsa:2006up,Lensky:2009uv}.

In precise predictions of neutron decay and inverse beta decay, the isovector axial-vector coupling constant of the nucleon \(g_A\) is one of the dominant sources of uncertainty. Chiral perturbation theory predicts significant corrections to \(g_A\) from both pion and \(\Delta\) loops~\cite{Bernard:2006te,Bernard:2025gto}, which are essential for understanding the quark-mass dependence of \(g_A\) and connecting lattice quantum chromodynamics (QCD) calculations to experimental measurements. A key question concerns the electromagnetic corrections to the relation between the physical \(g_A\) and the QCD-defined isovector axial-vector charge, \(g_A^{\text{QCD}}\), obtained from lattice-QCD calculations. As demonstrated in Refs.~\cite{Cirigliano:2022hob,Cirigliano:2022yyo,Cirigliano:2024nfi,Tomalak:2026wks}, the QED radiative corrections to the ratio \(g_A/(g_A^{\text{QCD}} g_V)\) from the matching to the pionless HBChPT can be large. \(\Delta\)-resonance contributions to this matching coefficient have not been previously evaluated and may significantly affect the interpretation of the lattice-QCD results.

In this work, we extend our previous analysis of pion-induced QED radiative corrections in neutron decay and IBD by including the \(\Delta(1232)\) resonance in HBChPT within the small-scale expansion power-counting scheme. We provide additional interactions involving the \(\Delta\) resonance and calculate the diagrams that appear when the \(\Delta\) is included as an explicit degree of freedom. We then compare the nucleon isovector axial-vector charge and IBD cross sections with and without \(\Delta\) contributions.

Our analysis reveals several important findings. First, we demonstrate that the \(\Delta\)-resonance contributions to the renormalization of the vector charge vanish identically, according to the Behrends-Sirlin-Ademollo-Gatto theorem~\cite{Behrends:1960nf,Ademollo:1964sr}. The corrections to the isovector axial-vector charge are of the same order as the pion contributions. Second, we show that including the \(\Delta\) degrees of freedom improves the convergence of the perturbation theory in the QED radiative correction to the experiment-over-lattice-QCD ratio \(g_A/(g_A^{\text{QCD}} g_V)\), reducing the size of the correction. Third, we find that the \(\Delta\) resonance increases the pion-induced QED radiative corrections to the IBD cross section by a factor of \(1.2\) to \(1.3\), with the relative contribution decreasing with antineutrino energy.

The paper is organized as follows. We specify kinematics, leading-order Lagrangian, and corresponding cross sections of the IBD reaction in Section~\ref{sec:leading_order}. In Section~\ref{sec:delta_lagrangian}, we present the leading and next-to-leading-order HBChPT Lagrangian for the $\Delta$ field, relevant for QED radiative corrections in neutron decay and IBD. Section~\ref{sec:delta_contributions_QED} provides QED radiative corrections from virtual diagrams with $\Delta$ and discusses associated bremsstrahlung. In subsection~\ref{subsec:delta_LO}, we present the leading-order QED radiative corrections. Corrections from the next-to-leading order HBChPT Lagrangian are provided in subsection~\ref{subsec:delta_NLO}. We present numerical results for the renormalization of the nucleon isovector vector and axial-vector charges and IBD cross sections in subsection~\ref{subsec:gV_gA_results} and subsection~\ref{subsec:IBD_results}, respectively, of Section~\ref{sec:results} and emphasize the differences between theories with and without explicit $\Delta$ degrees of freedom. Our conclusions and outlook are presented in Section~\ref{sec:conclusions}. The Wolfram Mathematica notebook and the Python library for fast and accurate evaluation of IBD cross sections are available at~\href{https://github.com/tomalak7/IBDxsec}{github.com/tomalak7/IBDxsec}.

\section{Charged-current semileptonic processes with nucleons at leading order} \label{sec:leading_order}

In this section, we specify the leading-order Lagrangian and the kinematics for charged-current semileptonic processes involving nucleons at low energies, considering the IBD reaction.

At energy scales below the pion mass, charged-current semileptonic processes with nucleons, neutron and inverse beta decay, are described by the effective Lagrangian $\mathcal L$~\cite{Ando:2004rk,Falkowski:2021vdg,Cirigliano:2022hob,Tomalak:2023xgm,Cirigliano:2023fnz,Cirigliano:2024nfi,Tomalak:2025jtn,Tomalak:2025okl}
\begin{equation} \label{eq:Lagrangian_at_leading_order}
	\mathcal L = - \sqrt{2} G_F V^\star_{ud} \overline{\nu}_{eL} \gamma_\rho e \cdot \overline{N}_v \left( g_V v^\rho - 2 g_A S^\rho \right) \tau^- N_v + \mathrm{h.c.},
\end{equation}
with the isovector vector, $g_V$, and the axial-vector, $g_A$, low-energy coupling constants (LECs), the scale-independent Fermi coupling constant $G_F$~\cite{Fermi:1934hr,Feynman:1958ty,vanRitbergen:1999fi,MuLan:2012sih}, and the Cabibbo-Kobayashi-Maskawa (CKM) quark mixing matrix element $V_{ud}$~\cite{Cabibbo:1963yz,Kobayashi:1973fv,ParticleDataGroup:2020ssz,Hardy:2020qwl}. In the above expression, $M$, $v^\mu = \left( 1, 0 \right)$, and $S^\mu$ ($v \cdot S = 0$) denote the mass, velocity, and spin of the heavy nucleon. The fields of proton $p$ and neutron $n$ are combined in the isodoublet $N_v = \left( p, n \right)^T$, while $\tau^- = \frac{\tau^1 - i \tau^2 }{2}$ stands for the lowering operator in the isospin space with the Pauli matrices $\tau^i$. The electron and neutrino fields are denoted as $e$ and ${\nu}_{eL}$, respectively.

In this paper, we evaluate $\Delta$-induced QED radiative corrections to IBD cross sections and renormalization of the isovector vector and axial-vector LECs, $g_V$ and $g_A$. To achieve these goals, it is sufficient to consider the IBD reaction $\overline{\nu}_e p \to e^+ n$.
\begin{figure}[ht]
\begin{center}
	\includegraphics[scale=1.46]{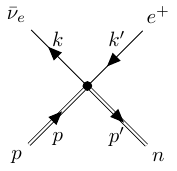}
	\caption{Kinematics in inverse beta decay. \label{fig:diagram}}
\end{center}
\end{figure}
We denote IBD kinematics in Fig.~\ref{fig:diagram}. The incoming antineutrino in the rest frame of the proton has momentum $k^\mu = (E_{\overline{\nu}_e},~\bk)$, the final positron has momentum $k^{\prime\mu} = (E_e,~\bk^\prime)$, the initial proton has momentum $p^\mu = (m_p,~0)$, with the proton mass $m_p$, and the final neutron momentum is determined by energy and momentum conservation as $p^{\prime\mu} = p^\mu + k^\mu - k^{\prime\mu}$. The momenta of external particles are constrained as $k^2 = 0$, $k^{\prime2} = m_e^2$, $p^{\prime2} = m_n^2$, with electron $m_e$ and neutron $m_n$ masses, respectively. The momentum transfer, $q^\mu$, is defined as $q^\mu = k^\mu - k^{\prime\mu}$.

For convenience, we also define two independent Lorentz-invariant Mandelstam variables. The squared momentum transfer $Q^2$,
\begin{equation} \label{eq:squared_momentum_transfer}
	Q^2 = - q^2 = -\left(p^\prime - p \right)^2 = m^2_p - m^2_n + 2 m_p \left( E_{\overline{\nu}_e} - E_e \right),
\end{equation}
and the squared energy in the center-of-mass reference frame $s$,
\begin{equation} \label{eq:squared_cmf_energy}
	s = \left(p + k \right)^2 = m^2_p + 2 m_p E_{\overline{\nu}_e}.
\end{equation}

Inverse beta decay is kinematically allowed only for antineutrino energies above the threshold $E_{\overline{\nu}_e} \ge E_{\overline{\nu}_e}^\mathrm{thr} = \frac{\left( m_n + m_e \right)^2}{2 m_p} - \frac{m_p}{2}$, with $E_{\overline{\nu}_e}^\mathrm{thr} \approx 1.806066~\mathrm{MeV}$. For a fixed antineutrino energy $E_{\overline{\nu}_e}$, the final positron energy is bounded in the narrow range $E_e^\mathrm{min} \le E_e \le E_e^\mathrm{max}$:
\begin{align}
	E_e^\mathrm{min} &= \frac{ \left( s - m^2_n + m^2_e \right) \left( m_p + E_{\overline{\nu}_e} \right) - E_{\overline{\nu}_e} \sqrt{\Sigma \left( s, m^2_n, m^2_e \right)} }{2 s}, \label{eq:positron_energy_min} \\
	E_e^\mathrm{max} &= \frac{ \left( s - m^2_n + m^2_e \right) \left( m_p + E_{\overline{\nu}_e} \right) + E_{\overline{\nu}_e} \sqrt{\Sigma \left( s, m^2_n, m^2_e \right)} }{2 s}, \label{eq:positron_energy_max}
\end{align}
with the kinematic triangle function $\Sigma \left( s, m^2_n, m^2_e \right) = \left( s - \left( m_n + m_e \right)^2 \right) \left( s - \left( m_n - m_e \right)^2 \right)$. Up to nucleon recoil corrections, the positron energy is related to the antineutrino energy as $E_e \approx E_{\overline{\nu}_e} + m_p - m_n + \mathcal{O} \left( \frac{1}{m_n} \right)$.

The differential IBD cross section at leading order can be evaluated from the interaction Lagrangian in Eq.~(\ref{eq:Lagrangian_at_leading_order}) as
\begin{equation} \label{eq:LO_static_diff}
	\frac{\mathrm{d}\sigma_{\rm LO}}{\mathrm{d} Q^2} = \frac{{G}_{F}^2 |V_{ud}|^2}{2\pi} \bigg[ \left( 1 - \frac{E_0}{E_{\overline{\nu}_e}} \right) \left( g_V^2 + g_A^2 \right) - \frac{Q^2+m^2_e}{4 E_{\overline{\nu}_e}^2} \left( g_V^2 - g_A^2 \right) \bigg],
\end{equation}
with $E_0 = \frac{ m^2_n + m^2_e - m^2_p}{2 m_n} \approx 1.292581~\mathrm{MeV}$, approximately determined by the neutron-proton mass difference: $E_0 \approx m_n - m_p$. All relevant corrections to Eq.~(\ref{eq:LO_static_diff}) are described in detail in Refs.~\cite{Tomalak:2025jtn,Tomalak:2025okl,Tomalak:2026btz}.

\section{$\Delta(1232)$ resonance in HBChPT} \label{sec:delta_lagrangian}

In this section, we provide details of the HBChPT Lagrangian with explicit $\Delta(1232)$ degrees of freedom, relevant to the study of $\Delta$-induced QED radiative corrections. We follow the notation and conventions of Refs.~\cite{Hemmert:1997ye,Siemens:2020vop} with the most recent fit results for the relevant low-energy coupling constants to the pion-nucleon scattering data, while we take the $\pi N$ interaction Lagrangian from Refs.~\cite{Gasser:1983yg,Meissner:1997ii,Muller:1999ww,Knecht:1999ag,Gasser:2002am,Gasser:1987rb,Krause:1990xc,Ecker:1995rk,Bernard:1995dp,Tomalak:2026btz}.

Four physical states of the $\Delta$ resonance $\left( \Delta^{++}, \Delta^+, \Delta^0, \Delta^- \right)$ form a triplet. They are conveniently combined in $3$ isospin doublets $\Psi^{i}_\mu$~\cite{Hemmert:1997ye}:
\begin{equation} \label{eq:delta_physical_fields}
	\Psi^1_\mu = \frac{1}{\sqrt{2}} \begin{pmatrix} \Delta^{++} - \frac{1}{\sqrt{3}} \Delta^0 \\ \frac{1}{\sqrt{3}} \Delta^0 - \Delta^- \end{pmatrix}_\mu, \qquad \Psi^2_\mu = \frac{i}{\sqrt{2}} \begin{pmatrix} \Delta^{++} + \frac{1}{\sqrt{3}} \Delta^0 \\ \frac{1}{\sqrt{3}} \Delta^+ + \Delta^- \end{pmatrix}_\mu, \qquad \Psi^3_\mu = - \sqrt{\frac{2}{3}} \begin{pmatrix} \Delta^{+} \\ \Delta^0 \end{pmatrix}_\mu,
\end{equation}
with an off-shell isospin constraint $\tau^i \Psi^i_\mu = 0$.

In this work, we consider a small-scale expansion of the heavy-baryon formulation of chiral perturbation theory~\cite{Hemmert:1997ye} and exploit spin-$3/2$ and isospin-$3/2$ projections onto light degrees of freedom $T^\mu_i$:\footnote{For consistency with the notation in~\cite{Tomalak:2025jtn,Tomalak:2025okl}, we use the neutron mass $m_n$ in this section.}
\begin{equation} \label{eq:light_delta_fields}
	T^\mu_i \left( x \right) = P_v^+ \xi_{3/2}^{i j} \left(\hat{P}^{3/2}_{33} \right)^{\mu \nu} \Psi^{j}_\nu \left( x \right) e^{i m_n v \cdot x},
\end{equation}
that can be obtained from the field $\Psi^{i}_\mu$ by acting with the spin projector $\left(\hat{P}^{3/2}_{33} \right)^{\mu \nu}$:
\begin{equation} \label{eq:spin_projector}
	\left(\hat{P}^{3/2}_{33} \right)^{\mu \nu} = g^{\mu \nu} - v^\mu v^\nu - \frac{4}{1-d} S^\mu S^\nu,
\end{equation}
in $d = 4 -2 \epsilon$ space-time dimensions, the isospin projector $\xi_{3/2}^{i j}$:
\begin{equation} \label{eq:isospin_projector}
	\xi_{3/2}^{i j} = \frac{2}{3} \delta^{i j} - \frac{i}{3} \varepsilon^{i j k} \tau^k,
\end{equation}
and the projector on light degrees of freedom $P_v^+$:
\begin{equation} \label{eq:light_fields_projector}
	P_v^+ = \frac{1 + \slashed{v}}{2}.
\end{equation}

The leading-order heavy-baryon $\pi \Delta$ interaction Lagrangian $\mathcal L^{\left( 1 \right)}_{\pi \Delta}$ is expressed in terms of the spin-$3/2$ isospin-$3/2$ projection onto light $\Delta$ degrees of freedom $T^\mu_i$ as~\cite{Hemmert:1997ye}
\begin{equation} \label{eq:leading_pi_delta_lagrangian}
	\mathcal L^{\left( 1 \right)}_{\pi \Delta} = - \overline{T}^\mu_i \left[ i v \cdot D^{i j } - \Delta_0 \delta^{i j} + g_1 S \cdot u^{i j}\right] g_{\mu \nu} T^\nu_j,
\end{equation}
with the $\Delta$-nucleon mass shift $\Delta_0 = m_\Delta - m_n \approx 293~\mathrm{MeV}$, the coupling constant $g_1$, the covariant derivative $D^{i j}_\mu$:
\begin{equation} \label{eq:covariant_derivative_with_isospin}
	D^{i j}_\mu = \delta^{i j} \partial_\mu + \Gamma_\mu^{ij},
\end{equation}
with the chiral connection $\Gamma_\mu^{ij}$
\begin{equation} \label{eq:chiral_connection_with_isospin}
	\Gamma_\mu^{ij} = \Gamma_\mu \delta^{i j} - i \varepsilon^{i j k} \langle \tau^k \Gamma_\mu \rangle,
\end{equation}
where the trace of any object ${\cal O}$ in the isospin space is $\langle {\cal O} \rangle$, expressed in terms of the pion degrees of freedom $\vec{\pi}$ entering $u$, $u^2 = U = e^{\frac{i \boldsymbol{\pi} \cdot \boldsymbol\tau}{F_\pi}}$, with the decay constant of the pion $F_\pi$, and external left and right sources $l_\mu$ and $r_\mu$, respectively, as
\begin{equation} \label{eq:chiral_connection}
	\Gamma_\mu = \frac{1}{2} \left[ u (\partial_\mu - i l_\mu) u^\dagger + u^\dagger (\partial_\mu - i r_\mu) u \right].
\end{equation}
$ u^{i j}_{\mu} = \xi^{i l}_{\frac{3}{2}} u_\mu \xi^{l j}_{\frac{3}{2}}$ denotes the isospin rotated pion field $u_\mu$:
\begin{equation} \label{eq:pion_field_umu}
	u_\mu = i \left[ u^\dagger (\partial_\mu - i r_\mu) u - u (\partial_\mu - i l_\mu) u^\dagger\right].
\end{equation}
To evaluate low-energy charged-current processes with leptons and nucleons, we account for the electromagnetic and electroweak interactions through the external sources as
\begin{align}
    l_\mu &= - e Q A_\mu + Q_W \, \overline{e} \gamma_\mu \nu_{eL} + Q_W^\dagger \overline{\nu}_{eL} \gamma_\mu e, \\
    r_\mu &= - e Q A_\mu, \label{eq:sources}
\end{align}
with the photon field $A_\mu$, the proton charge $e$, the charge matrix $Q = {\rm diag} (1, 0)$, and the electroweak matrix $Q_W^\dagger = -2 \sqrt{2} {G}_{F} V_{ud}^\star \tau^-$.

At leading order, the $\pi N \Delta$ interaction Lagrangian $\mathcal L^{\left( 1 \right)}_{\pi N \Delta} $ can be written as~\cite{Hemmert:1997ye}
\begin{equation} \label{eq:leading_piN_delta_lagrangian}
	\mathcal L^{\left( 1 \right)}_{\pi N \Delta} = h \overline{T}^\mu_i w_\mu^i N + \mathrm{h.c.},
\end{equation}
with the $\pi N \Delta$ coupling constant $h$ and $w^i_\mu = \frac{1}{2} \langle \tau^i u_\mu \rangle$.

At next-to-leading order, the relevant terms of the $\pi N \Delta$ interaction Lagrangian $\mathcal L^{\left( 2 \right)}_{\pi N \Delta} $ can be written as
 \begin{align} \label{eq:next_to_leading_piN_delta_lagrangian}
	\mathcal L^{\left( 2 \right)}_{\pi N \Delta} &= \overline{T}^\mu_i \left[ \frac{i b_1 f^+_{\mu \nu} S^\nu}{2 m_n} + \left( b_3 + b_6 \right) i w_{\mu \nu}^i v^\nu + b_4 w^i_\mu S \cdot u + b_5 u_\mu S \cdot w^i \right] N + \mathrm{h.c.} \nonumber \\
	&- \frac{1}{2 m_n} \overline{T}_\mu^i \left[ \frac{2}{d - 1} h g_1 z_0 u^\mu_{i j} \xi^{j k} S \cdot w^k + 2 h i D^\mu_{ij} \xi^{jk} v \cdot w^k \right] N + \mathrm{h.c.},
\end{align}
with the NLO LECs $b_1$-$b_6$, the tensor $w^i_{\mu \nu} = \frac{1}{2} \langle \tau^i \left[ D_\mu, u_\nu \right] \rangle$, with $D_\mu = \partial_\mu + \Gamma_\mu$, $f^+_{\mu\nu} =  u^\dagger \left(  \partial_\mu r_\nu -  \partial_\nu r_\mu - i \left[ r_\mu, r_\nu \right] \right) u + u \left(  \partial_\mu l_\nu -  \partial_\nu l_\mu - i \left[ l_\mu, l_\nu \right] \right) u^\dagger$, and the off-shell parameter $z_0$.\footnote{The coupling constant $b_1 = 3.61$ is presented in Refs.~\cite{Griesshammer:2012we,McGovern:2012ew}.} To restore the explicit $d$-dependence in the Lagrangian, an additional contribution to the pion-nucleon interaction Lagrangian $\mathcal L^{\left( 2 \right)}_{\pi N}$ is added~\cite{Hemmert:1997ye},
\begin{align} \label{eq:next_to_leading_piN_lagrangian}
	\mathcal L^{\left( 2 \right)}_{\pi N} &= - \frac{h^2}{2 m_n} \frac{4}{d-1} \left( 2 z_0 + \left( d - 1 \right) z_0^2 \right) \overline{N} S \cdot w^i \xi^{ij} S \cdot w^j N \nonumber \\
	&- \frac{h^2}{2 m_n} \frac{1}{d-1} \left( 4 \left( d - 2 \right) + 2 \left( d - 3 \right) z_0 - z_0^2 \right) \overline{N} v \cdot w^{i} \xi^{ij} v \cdot w^{j} N.
\end{align}
With appropriate redefinitions of LECs, the redundant dependence on LECs $b_3$ and $b_4$ and the off-shell parameter $z_0$ disappears~\cite{Krebs:2009bf,Siemens:2020vop}.

Parts of recoil corrections and $\Delta$ contributions are generated by the following terms of the next-to-leading order $\pi \Delta$ interaction Lagrangian $\mathcal L^{\left( 2 \right)}_{\pi \Delta}$:
 \begin{align} \label{eq:next_to_leading_pi_delta_lagrangian}
	\mathcal L^{\left( 2 \right)}_{\pi \Delta} &= \frac{1}{2 m_n} \overline{T}_i^\mu \left[ D^{ik}_\alpha D^{kj}_\beta g^{\alpha \beta} - v \cdot D^{ik} v \cdot D^{k j} - \left[ S^\alpha, S^\beta \right] \left( D^{ik}_\alpha D^{kj}_\beta - D^{ik}_\beta D^{kj}_\alpha \right) \right] g_{\mu \nu} T^\nu_j \nonumber \\
	&+ \frac{i g_1}{2 m_n} \overline{T}_i^\mu \left( S \cdot D^{i k} v \cdot u^{k j} + v \cdot u^{i k} S \cdot D^{k j} \right) g_{\mu \nu} T^\nu_j - \overline{T}_i^\mu c_1^\Delta m^2_\pi < U^\dagger + U> \delta^{ij} g_{\mu \nu} T^\nu_j,
\end{align}
where we present only the terms relevant to this work.

\section{$\Delta$-induced QED radiative corrections} \label{sec:delta_contributions_QED}

In this section, we evaluate QED radiative corrections to IBD cross sections and renormalization of the isovector vector $g_V$ and axial-vector $g_A$ LECs induced by $\Delta$ degrees of freedom at leading order in the electromagnetic coupling constant $\alpha$. We evaluate contributions from leading and next-to-leading HBChPT interactions in subsections~\ref{subsec:delta_LO} and~\ref{subsec:delta_NLO}, respectively. We do not provide known results in the $\Delta$-less theory~\cite{Cirigliano:2022hob,Tomalak:2026wks,Tomalak:2026btz} but emphasize the difference in LECs between theories with vs without explicit $\Delta$ degrees of freedom.

\subsection{Leading-order HBChPT contributions} \label{subsec:delta_LO}

Repeating the analysis of pionless diagrams in Refs.~\cite{Tomalak:2021lif,Tomalak:2026btz}, we find that the sum of diagrams with virtual photon, electron, nucleon, $\Delta$, and pion fields at LO, but without the isospin-breaking interaction, is suppressed by the electron mass and, consequently, we neglect all such diagrams, both in forward and non-forward kinematics. The same argument also applies to the $\Delta$-induced bremsstrahlung at the amplitude level.

\begin{figure}[ht]
\begin{center}
	\includegraphics[scale=0.316]{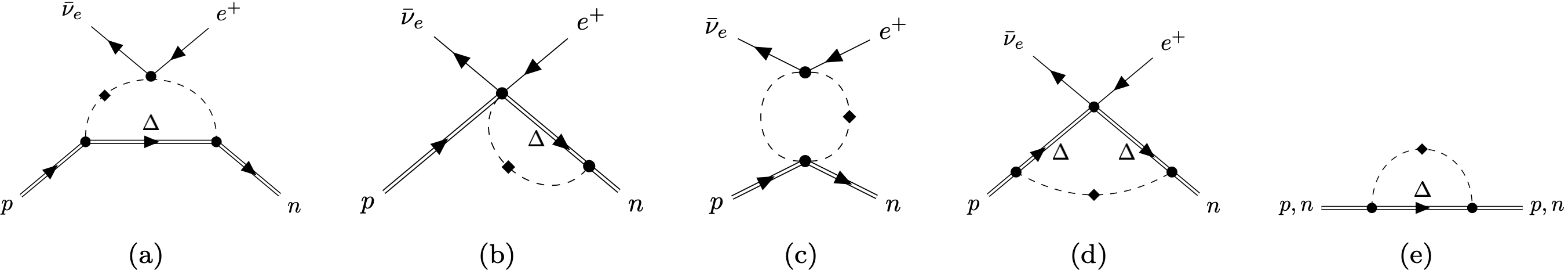}
	\caption{$\Delta$-induced QED radiative corrections without virtual photons at leading and next-to-leading order. The pion isospin-breaking vertex is shown as a square, with all possible placements of this vertex implied. The nucleon field-renormalization diagram (e) is also illustrated.\label{fig:leading_order_Zpi_diagrams_delta}}
\end{center}
\end{figure}

The remaining QED radiative corrections at LO are induced by the isospin-breaking LEC $Z_\pi \approx 0.81$, which generates the diagrams in figure~\ref{fig:leading_order_Zpi_diagrams_delta} and also gives rise to $\Delta$-induced contributions to the proton and neutron field renormalization factors.

In the following, we provide the results for each diagram in figure~\ref{fig:leading_order_Zpi_diagrams_delta}.\footnote{We use dimensional regularization with the chiral version of the modified minimal subtraction scheme~\cite{Gasser:1983yg} and subtract the following expression with the ultraviolet pole $\frac{1}{\varepsilon}$
\begin{equation}
	\frac{1}{\varepsilon} - \gamma_E + \ln \left( 4 \pi \right) + 1.
\end{equation}
In the evaluation, we have reproduced the master integrals from Refs.~\cite{Cho:1992cf,Falk:1993fr,Boyd:1994pa,Stewart:1998ke,Bouzas:1999ug,Fajfer:2001ad,Bouzas:2001py,Davydychev:2001ui,Zupan:2002je,Bouzas:2002xi,Becirevic:2002sc}.}

a) The contribution of this diagram can be represented as the kinematic-dependent change of the vector coupling constant
\begin{align} \label{eq:LO_a_gV}
	\delta g_V &= \frac{2}{3} \frac{\alpha}{\pi} h^2 Z_\pi \left( \ln \frac{\mu^2}{m^2_\pi} + \frac{1}{\beta_\Delta} \ln \frac{1-\beta_\Delta}{1+\beta_\Delta} - 2 \right) + \frac{2}{9} \frac{\alpha}{\pi} h^2 Z_\pi \left( \frac{c^\mathrm{LO}_\beta}{\beta} \ln \frac{\beta - 1}{\beta + 1} - \frac{c^\mathrm{LO}_{\beta_\Delta}}{\beta_\Delta} \ln \frac{1-\beta_\Delta}{1+\beta_\Delta} \right) \nonumber \\
	&+ \frac{2}{9} \frac{\alpha}{\pi} h^2 Z_\pi \left( 4 + \frac{\Sigma}{\gamma} \right),
\end{align}
with $\beta_\Delta = \sqrt{1- \frac{m_\pi^2}{\Delta_0^2}}$, $\beta = \sqrt{1+\frac{4m^2_\pi}{Q^2}}$, $\gamma = \frac{Q}{2 \Delta_0}$, the sum $\Sigma$,
\begin{equation} \label{eq:sigma}
	\Sigma = \sum \limits_{\sigma_1, \sigma_2 = \pm 1} \sigma_1 \left( \mathrm{Li}_2 \frac{1-\sigma_1 \gamma}{1 - \sigma_2 \beta \gamma} + \mathrm{Li}_2 \frac{ 1 + \frac{\sigma_1}{\gamma}}{1 + \frac{\sigma_2 \beta_\Delta}{ \gamma}} \right),
\end{equation}
and the coefficients $c^\mathrm{LO}_{\beta}$ and $c^\mathrm{LO}_{\beta_\Delta}$ defined as
\begin{align} \label{eq:coefficients_LO}
	c^\mathrm{LO}_\beta &= 1+ 2 \beta^2 - \frac{1}{\beta^2_\Delta - \gamma^2}, \\
	c^\mathrm{LO}_{\beta_\Delta} &= 2 - \frac{\gamma^2}{\beta^2_\Delta - \gamma^2},
\end{align}
and a non-factorizable correction to the IBD cross section
\begin{equation} \label{eq:LO_NF_a}
	\frac{\mathrm{d}\sigma}{\mathrm{d} Q^2} = \frac{2}{9} \frac{\alpha}{\pi} h^2 Z_\pi \Sigma \frac{{G}_{F}^2 |V_{ud}|^2}{2\pi} \frac{Q}{E_{\overline{\nu}_e}} g_A.
\end{equation}

b) Neglecting terms suppressed by the electron mass and recoil $\frac{\alpha}{\pi} \frac{E_{\overline{\nu}_e}}{m_n}$, this diagram does not contribute to charged-current semileptonic processes with a nucleon.

c) This diagram is absent at LO HBChPT.

d) This diagram contributes only to the renormalization of the isovector vector and axial-vector LECs as
\begin{align}
	\delta g_V &= - 2 \frac{\alpha}{\pi} h^2 Z_\pi \left( \ln \frac{\mu^2}{m^2_\pi} + \frac{1}{\beta_\Delta} \ln \frac{1-\beta_\Delta}{1+\beta_\Delta} - 2 \right), \label{eq:LO_c_gV} \\
	 \frac{\delta g_A}{g_A^{(0)}} &= - \frac{10}{27} \frac{\alpha}{\pi} \frac{g_1}{g_A^{(0)}} h^2 Z_\pi \left( \ln \frac{\mu^2}{m^2_\pi} + \frac{1}{\beta_\Delta} \ln \frac{1-\beta_\Delta}{1+\beta_\Delta} - \frac{46}{15}\right), \label{eq:LO_c_gA}
\end{align}
with the nucleon isovector axial-vector coupling constant in the chiral limit $g_A^{(0)}$.

d$^\star$) In addition to the contributions with a $\Delta$ inside the virtual loop, the $\pi N \Delta$ interactions in Eq.~(\ref{eq:leading_piN_delta_lagrangian}) generate a diagram in topology (d) containing one nucleon and one $\Delta$ propagator. This diagram contributes to the isovector axial-vector LEC as
\begin{equation}
	 \frac{\delta g_A}{g_A^{(0)}} = - \frac{4}{3}  \frac{\alpha}{\pi} h^2 Z_\pi \left(  \ln \frac{\mu^2}{m^2_\pi} + \beta_\Delta \ln \frac{1-\beta_\Delta}{1+\beta_\Delta} + \frac{2}{3} - \frac{\pi m_\pi}{\Delta_0}\right). \label{eq:LO_c2_gA}
\end{equation}

e) The nucleon field renormalization factors renormalize the isovector vector and axial-vector LECs as follows:
 \begin{equation} \label{eq:ZN_LO}
	\delta g_V = \frac{\delta g_A}{g_A^{(0)}} = \frac{4}{3} \frac{\alpha}{\pi} h^2 Z_\pi \left( \ln \frac{\mu^2}{m^2_\pi} + \frac{1}{\beta_\Delta} \ln \frac{1-\beta_\Delta}{1+\beta_\Delta} - 2 \right).
\end{equation}

Combining all pieces, we obtain the $\Delta$-induced renormalization of the isovector vector and axial-vector LECs at leading order
\begin{align}
	\delta g^\mathrm{LO}_V &= 0,  \label{eq:renormalization_gV_LO} \\
    \frac{\delta g^\mathrm{LO}_A}{g_A^{(0)}} &= \frac{4}{3} \frac{\alpha}{\pi} h^2 Z_\pi \left[ \left( 1 - \frac{5}{18} \frac{g_1}{g_A^{(0)}} \right) \left( \ln \frac{\mu^2}{m^2_\pi} + \frac{1}{\beta_\Delta} \ln \frac{1-\beta_\Delta}{1+\beta_\Delta} - 2 \right) + \frac{8}{27} \frac{g_1}{g_A^{(0)}} \right]\nonumber \\
	 &- \frac{4}{3}  \frac{\alpha}{\pi} h^2 Z_\pi \left( \ln \frac{\mu^2}{m^2_\pi} + \beta_\Delta \ln \frac{1-\beta_\Delta}{1+\beta_\Delta} + \frac{2}{3} - \frac{\pi m_\pi}{\Delta_0}\right). \label{eq:renormalization_gA_LO}
\end{align}
In agreement with the Behrends-Sirlin-Ademollo-Gatto theorem~\cite{Behrends:1960nf,Ademollo:1964sr}, the vector coupling constant $g_V$ does not get renormalized. Contributions in Eqs.~(\ref{eq:renormalization_gV_LO}) and~(\ref{eq:renormalization_gA_LO}) are included in the LECs $g_V$ and $g_A$ in the calculation of the effective field theory without pions. The remaining kinematic dependence of the LO $\Delta$-induced QED radiative corrections can be expressed as the kinematic-dependent part of the vector coupling constant shift in Eq.~(\ref{eq:LO_a_gV}) and the non-factorizable IBD cross section in Eq.~(\ref{eq:LO_NF_a}).

\subsection{Next-to-leading-order HBChPT contributions} \label{subsec:delta_NLO}

As for leading-order corrections, the next-to-leading-order contribution from the sum of diagrams with virtual photon, electron, nucleon, $\Delta$ fields, and pions, but without isospin-breaking interaction and $b_1$ coupling, in virtual and real radiative corrections is suppressed by the electron mass and can be neglected. In contrast to the pion-induced QED contribution at NLO~\cite{Tomalak:2026btz}, the $\Delta$ resonance does not generate a diagram with a five-point, pionless, four-fermion-photon vertex at NLO. Therefore, neither corrections of order $\frac{\alpha}{\pi}\frac{E_{\overline{\nu}_e}}{m_n}$ to IBD nor corrections of order $\frac{\alpha}{\pi}\frac{\Delta_0}{m_n}$ to the renormalization of LECs are generated by NLO interactions at $5$.

All $\Delta$-induced NLO QED radiative corrections are described by diagrams in figure~\ref{fig:leading_order_Zpi_diagrams_delta}, with recoil or NLO vertices on the baryon side, and the pionless triangle diagrams with the photon from the magnetic dipole $N\to\Delta$ transition entering with the LEC $b_1$. All contributions from the pion isospin-breaking interaction are manifestly gauge-invariant and do not generate any bremsstrahlung at the order of the calculation in this paper, both at LO and NLO HBChPT. Consequently, all radiation from the LO HBChPT interactions and $\Delta$-induced radiation at NLO, excluding contributions from $b_1$, is suppressed by the electron mass and therefore does not require precise calculations. Neglecting the renormalization of the nucleon magnetic moments and higher-order operators, as well as the electron-mass-suppressed terms, we provide the NLO contributions from each topology in figure~\ref{fig:leading_order_Zpi_diagrams_delta} as follows.

a) The contribution of this diagram, after removing the renormalization of the nucleon magnetic moments and higher-order operators, can be represented as the kinematic-dependent change of the vector coupling constant
\begin{align} \label{eq:NLO_a_gV}
	\delta g_V &= 2 \frac{\alpha}{\pi} \frac{\Delta_0}{m_n} h^2 Z_\pi \left( \ln \frac{\mu^2}{m^2_\pi} + \frac{1 + 5 \beta^2_\Delta}{6 \beta_\Delta} \ln \frac{1-\beta_\Delta}{1+\beta_\Delta} - 1 \right) + \frac{2}{9} \frac{\alpha}{\pi} \frac{\Delta_0}{m_n} h^2 Z_\pi \left( \beta^2_\Delta - \gamma^2 \right) \frac{\Sigma}{\gamma} \nonumber \\
	&+ \frac{2}{9} \frac{\alpha}{\pi} \frac{\Delta_0}{m_n} h^2 Z_\pi \left( \frac{c^\mathrm{NLO}_\beta}{\beta} \ln \frac{\beta - 1}{\beta + 1} - \frac{c^\mathrm{NLO}_{\beta_\Delta}}{\beta_\Delta} \ln \frac{1-\beta_\Delta}{1+\beta_\Delta} + 5 - \frac{\beta^2_\Delta}{\beta^2_\Delta - \gamma^2} \right),
\end{align}
with the coefficients $c^\mathrm{NLO}_{\beta}$ and $c^\mathrm{NLO}_{\beta_\Delta}$ defined as
\begin{align} \label{eq:coefficients_NLO}
	c^\mathrm{NLO}_\beta &= \frac{ 2 \beta^2 \beta^2_\Delta - \gamma^2}{\beta^2_\Delta - \gamma^2} + \frac{\gamma^2}{\left(\beta^2_\Delta - \gamma^2 \right)^2}, \\
	c^\mathrm{NLO}_{\beta_\Delta} &= 2 \beta^2_\Delta - \frac{1}{2}\frac{\beta^2_\Delta \gamma^2}{\beta^2_\Delta - \gamma^2} + \frac{\beta^2_\Delta \gamma^2}{\left( \beta^2_\Delta - \gamma^2 \right)^2} + \frac{3}{2} \frac{ \gamma^2}{\beta^2_\Delta - \gamma^2},
\end{align}
and a non-factorizable correction to the IBD cross section
\begin{align} \label{eq:NLO_NF_a}
	\frac{\mathrm{d}\sigma}{\mathrm{d} Q^2} &= \frac{1}{9} \frac{\alpha}{\pi} \frac{Q}{m_n} h^2 Z_\pi \frac{\gamma^2}{\beta^2_\Delta - \gamma^2} \left(\beta \ln \frac{\beta - 1}{\beta + 1} - \frac{1}{\beta_\Delta} \ln \frac{1-\beta_\Delta}{1+\beta_\Delta}\right) \frac{{G}_{F}^2 |V_{ud}|^2}{2\pi} \frac{Q}{E_{\overline{\nu}_e}} g_A.
\end{align}

b) Recoil corrections renormalize the vector coupling constant as
\begin{align} \label{eq:NLO_b_gV}
	\delta g_V &= - \frac{4}{3} \frac{\alpha}{\pi} \frac{\Delta_0}{m_n} h^2 Z_\pi \left( \ln \frac{\mu^2}{m^2_\pi} + \beta_\Delta \ln \frac{1-\beta_\Delta}{1+\beta_\Delta}\right).
\end{align}

NLO HBChPT interactions contribute to the renormalization of the nucleon isovector axial-vector coupling constant at next-to-leading order as
\begin{equation} \label{eq:NLO_b_gA}
	\frac{\delta g_A}{g_A^{(0)}} = 2 \frac{\alpha}{\pi} h \Delta_0 \left( \frac{17}{9} b_5 + 2 b_4 \right) Z_\pi \left( \ln \frac{\mu^2}{m^2_\pi} + \beta_\Delta \ln \frac{1-\beta_\Delta}{1+\beta_\Delta} + \frac{2}{9} \right) + \frac{28}{81} \frac{\alpha}{\pi} h \Delta_0 b_5 Z_\pi.
\end{equation}

Neglecting the renormalization of the nucleon magnetic moments and higher-order operators, this diagram does not produce a kinematic-dependent non-factorizable contribution to the IBD cross section.

b$^\star$) In addition to contributions with $\Delta$ inside the virtual loop, contact nucleon-nucleon interactions in Eq.~(\ref{eq:next_to_leading_piN_lagrangian}) generate the diagram in topology (b) with a nucleon propagator. However, the part of this diagram that is independent of the off-shell parameter $z_0$ vanishes in both forward and non-forward kinematics, while the dependence on $z_0$ is removed by the shift in NLO LECs~\cite{Siemens:2020vop},
\begin{align}
    b_4 &\to b_4 - \frac{2}{3} \frac{g_1}{m_n} h \frac{1}{d-1} z_0, \\
    b_5 &\to b_5 + \frac{g_1}{m_n} h \frac{1}{d-1}  z_0.
\end{align}

c) This diagram contributes only the terms that depend on the off-shell parameter $z_0$. These terms are removed by the shift in the NLO LEC $c_4$~\cite{Siemens:2020vop}
\begin{align}
    c_4 &\to c_4 - \frac{1}{3} \frac{h^2}{m_n} \frac{1}{d-1} \left( 2 z_0 + \left( d - 1 \right) z^2_0 \right).   
\end{align}

d) The contribution of this diagram can be represented as the kinematic-dependent change of the isovector vector and axial-vector coupling constants
\begin{align} \label{eq:NLO_c_gV}
	\delta g_V &= - 2 \frac{\alpha}{\pi} \frac{\Delta_0}{m_n} h^2 Z_\pi \left( \ln \frac{\mu^2}{m^2_\pi} + \frac{1 + \beta^2_\Delta}{2 \beta_\Delta} \ln \frac{1-\beta_\Delta}{1+\beta_\Delta} - 3 \right) \nonumber \\
	&+ \frac{\alpha}{\pi} \frac{Q^2}{m_n \Delta_0} h^2 Z_\pi \frac{1}{\beta^2_\Delta} \left( 1 + \frac{1 - \beta^2_\Delta}{2 \beta_\Delta} \ln \frac{1-\beta_\Delta}{1+\beta_\Delta}\right), \\
	 \frac{\delta g_A}{g_A^{(0)}} &= - \frac{10}{27} \frac{\alpha}{\pi} \frac{g_1}{g_A^{(0)}} \frac{\Delta_0}{m_n} h^2 Z_\pi \left( \ln \frac{\mu^2}{m^2_\pi} + \frac{1 + \beta^2_\Delta}{2 \beta_\Delta} \ln \frac{1-\beta_\Delta}{1+\beta_\Delta} - \frac{37}{3} \right) \nonumber \\
	&+ \frac{5}{27} \frac{\alpha}{\pi} \frac{g_1}{g_A^{(0)}} \frac{Q^2}{m_n \Delta_0} h^2 Z_\pi \frac{1}{\beta^2_\Delta} \left( 1 + \frac{1 - \beta^2_\Delta}{2 \beta_\Delta} \ln \frac{1-\beta_\Delta}{1+\beta_\Delta}\right). \label{eq:NLO_c_gA}
\end{align}

d$^\star$) In addition to contributions with $\Delta$ inside the virtual loop, the diagram in topology (d) with one nucleon and one $\Delta$ propagator is also present. This diagram contributes to the kinematic-dependent change of the isovector axial-vector LEC
\begin{equation}
	 \frac{\delta g_A}{g_A^{(0)}} = -  \frac{80}{27} \frac{\alpha}{\pi} \frac{\Delta_0}{m_n} h^2 Z_\pi \left( \ln \frac{\mu^2}{m^2_\pi} + \beta_\Delta \ln \frac{1-\beta_\Delta}{1+\beta_\Delta} + \frac{13}{60} \right) - \frac{1}{3} \frac{\alpha}{\pi} \frac{Q^2}{m_n \Delta_0} h^2 Z_\pi \frac{1}{\beta_\Delta} \ln \frac{1-\beta_\Delta}{1+\beta_\Delta}. \label{eq:NLO_c2_gA}
\end{equation}

e) NLO HBChPT interactions contribute to the proton and neutron field renormalization factors as recoil corrections
\begin{equation} \label{eq:Zp_NLO}
	\delta g_V = \frac{\delta g_A}{g_A^{(0)}} = \frac{4}{3} \frac{\alpha}{\pi} \frac{\Delta_0}{m_n} h^2 Z_\pi \left( \ln \frac{\mu^2}{m^2_\pi} + \frac{1 + \beta^2_\Delta}{2 \beta_\Delta} \ln \frac{1-\beta_\Delta}{1+\beta_\Delta} - 3 \right).
\end{equation}
The results for the nucleon field renormalization factors can be readily applied to all low-energy processes involving the nucleon.

Additionally, the pionless triangle diagrams with the magnetic dipole $N \to \Delta$ transition contribute to the isovector axial-vector coupling constant as
\begin{equation} \label{eq:NLO_triangle_gA}
	\frac{\delta g_A}{g_A^{(0)}} = \frac{1}{9} \frac{\alpha}{\pi} h b_1 \frac{\Delta_0}{m_n} \left( \ln \frac{\mu^2}{4 \Delta_0^2} - \frac{4}{3} \right),
\end{equation}
up to corrections suppressed by the positron mass and/or energy. After subtracting the decoupling-breaking terms~\cite{Siemens:2017opr,Siemens:2020vop} that diverge in the limit $\Delta_0 \to \infty$, the contribution of the triangle diagrams vanishes.

Combining all pieces, we obtain the renormalization of the isovector vector and axial-vector coupling constants at next-to-leading order
\begin{align} \label{eq:renormalization_gV_NLO}
	\delta g^\mathrm{NLO}_V &= 0, \\
    \frac{\delta g^\mathrm{NLO}_A}{g_A^{(0)}} &= \frac{4}{3} \frac{\alpha}{\pi} \frac{\Delta_0}{m_n} h^2 Z_\pi \left[ \left( 1 - \frac{5}{18} \frac{g_1}{g_A^{(0)}} \right) \left( \ln \frac{\mu^2}{m^2_\pi} + \frac{1 + \beta^2_\Delta}{2 \beta_\Delta} \ln \frac{1-\beta_\Delta}{1+\beta_\Delta} - 3 \right) + \frac{70}{27} \frac{g_1}{g_A^{(0)}} - \frac{13}{27} \right] \nonumber \\
    &+ 2 \frac{\alpha}{\pi} h \Delta_0 \left( 2 b_4 + \frac{17}{9} b_5 - \frac{40}{27} \frac{h}{m_n} \right) Z_\pi \left( \ln \frac{\mu^2}{m^2_\pi} + \beta_\Delta \ln \frac{1-\beta_\Delta}{1+\beta_\Delta} + \frac{2}{9} \right) \nonumber \\
    &+ \frac{1}{9} \frac{\alpha}{\pi} h b_1 \frac{\Delta_0}{m_n} \left( \ln \frac{\mu^2}{4 \Delta_0^2} - \frac{4}{3} \right) + \frac{28}{81} \frac{\alpha}{\pi} h \Delta_0 b_5 Z_\pi. \label{eq:renormalization_gA_NLO}
\end{align}
In agreement with the Behrends-Sirlin-Ademollo-Gatto theorem~\cite{Behrends:1960nf,Ademollo:1964sr}, the vector coupling constant $g_V$ does not renormalize, with the vanishing result separately for recoil corrections to the $\Delta$ propagator from Eq.~(\ref{eq:next_to_leading_pi_delta_lagrangian}), recoil corrections to the interactions in the second line in Eq.~(\ref{eq:next_to_leading_piN_delta_lagrangian}), as well as NLO contributions from the Wilson coefficients $c_1^\Delta$, $b_1$-$b_6$. By directly evaluating contributions from the Lagrangian in Eq.~(\ref{eq:next_to_leading_piN_delta_lagrangian}), we find that the redundant $\pi N$ scattering NLO Wilson coefficients $b_3$ and $b_6$~\cite{Siemens:2020vop} do not contribute at NLO. The dependence on the off-shell parameter $z_0$ is removed by the shift in NLO LECs~\cite{Siemens:2020vop},
\begin{align}
    c_3 &\to c_3 - \frac{1}{3} \frac{1}{m_n} h^2 \frac{1}{d-1} \left( 2 z_0 + \left( d - 1 \right) z^2_0 \right), \\
    c_4 &\to c_4 - \frac{1}{3} \frac{1}{m_n} h^2 \frac{1}{d-1} \left( 2 z_0 + \left( d - 1 \right) z^2_0 \right), \\
    b_4 &\to b_4 - \frac{2}{3} \frac{g_1}{m_n} h \frac{1}{d-1} z_0, \\
    b_5 &\to b_5 + \frac{g_1}{m_n} h \frac{1}{d-1} z_0,
\end{align}
in agreement with the independence of matrix elements from the off-shell parameter $z_0$~\cite{Krebs:2009bf,Siemens:2020vop}. Contributions in Eqs.~(\ref{eq:renormalization_gV_NLO}) and~(\ref{eq:renormalization_gA_NLO}) are included in the LECs $g_V$ and $g_A$ in the calculation of the effective field theory without pions. The remaining kinematic dependence of the NLO $\Delta$-induced QED radiative corrections can be expressed as the kinematic-dependent part of the isovector vector and axial-vector coupling constant shifts in Eqs.~(\ref{eq:NLO_a_gV}),~(\ref{eq:NLO_c_gV}), (\ref{eq:NLO_c_gA}), and~(\ref{eq:NLO_c2_gA}) and the non-factorizable IBD cross section in Eq.~(\ref{eq:NLO_NF_a}). The $\Delta$-induced QED radiative corrections to the kinematic dependence of IBD cross sections are independent of the NLO HBChPT LECs.

\section{Results and Discussion} \label{sec:results}

In this section, we present results for the pion- and $\Delta$-induced QED radiative corrections at LO and NLO HBChPT. In section~\ref{subsec:gV_gA_results}, we study the renormalization of the isovector vector and the axial-vector coupling constants. In section~\ref{subsec:IBD_results}, we provide kinematic-dependent QED radiative corrections to IBD cross sections. For all results in section~\ref{subsec:gV_gA_results}, we subtract the decoupling-breaking terms~\cite{Siemens:2017opr,Siemens:2020vop,Gialidi:2026yyk} from expressions in section~\ref{sec:delta_contributions_QED}.

\subsection{Renormalization of $g_V$ and $g_A$} \label{subsec:gV_gA_results}

The vector coupling constant $g_V$ does not renormalize at leading and next-to-leading orders, according to the Behrends-Sirlin-Ademollo-Gatto theorem~\cite{Behrends:1960nf,Ademollo:1964sr}. The contribution of the $\Delta$ resonance to the renormalization of the isovector axial-vector coupling constant depends on the LECs $h$ and $g_1$ at LO and, in addition, on the Wilson coefficients $b_4$ and $b_5$ at NLO. In the following table~\ref{tab:gA_renormalization}, we compare the pion-induced QED radiative corrections to the nucleon isovector axial-vector charge to the contribution from loops with $\Delta$ and provide the resulting correction to $\frac{g_A}{g^\mathrm{QCD}_A g_V}$~\cite{Tomalak:2026wks}, for comparison with lattice-QCD determinations of the nucleon isovector axial-vector charge $g^\mathrm{QCD}_A$. We exploit an exact cancellation pattern observed between the HBChPT radiative corrections and the matching coefficient to the Standard Model~\cite{Cirigliano:2024nfi}, evaluate the results of this paper for the $\Delta$-induced infrared parts of the nucleon matrix elements, and take estimates of the nucleon contribution to these matrix elements from Ref.~\cite{Tomalak:2026wks}. To obtain numerical results, we exploit fits to $\pi N$ scattering data in $\Delta$-less theory from Ref.~\cite{Siemens:2017opr}, with subthreshold parameters constrained by Roy-Steiner analysis, and fits in the theory with $\Delta$ from Ref.~\cite{Siemens:2020vop}, for the complex mass approach. We investigate the convergence of the HBChPT by presenting the results for fits at $\mathcal{O} \left( p^2 \right)$, $\mathcal{O} \left( p^3 \right)$, and $\mathcal{O} \left( p^4 \right)$. For $\mathcal{O} \left( p^3 \right)$, we take $b_4 = b_5 = 0$ and provide only the recoil NLO contribution.
\begin{table}
	\centering
	\caption{Pion- and $\Delta$-induced QED radiative corrections to $\frac{g_A}{g^\mathrm{QCD}_A g_V} - 1$ are presented. Results are evaluated for HBChPT Wilson coefficients obtained from fits in Refs.~\cite{Siemens:2017opr,Siemens:2020vop} at $\mathcal{O} \left( p^2 \right)$, $\mathcal{O} \left( p^3 \right)$, and $\mathcal{O} \left( p^4 \right)$, employing the complex mass approach, and in the $\Delta$-less theory. For $\mathcal{O} \left( p^3 \right)$, only recoil NLO contributions from the $\Delta$ resonance are included.}
	\label{tab:gA_renormalization}
    \vspace{0.25cm}
	\begin{tabular}{|c|c|c|c|c|c|c|c|c|c|}
	\hline
	$\%$ 	& $\Delta$-less, $p^2$ & $\Delta$-less, $p^3$ & $\Delta$-less, $p^4$ & $p^2$ & $p^3$ & $p^4$ \\
	\hline
	$\pi$ LO		& 0.99(0.18) & 0.99(0.18) & 0.99(0.18) & 0.99(0.18) & 0.99(0.18) & 0.99(0.18) \\
	$\Delta$ LO	    & - & - & - & - & 0.3 & 0.4 \\
	$\pi$ NLO		& 2.0 & 1.9 & 1.4 & 0.6 & 1.0 & 0.3 \\
	$\Delta$ NLO	& - & - & - & - & -0.1 & 0.7 \\ \hline
	other~\cite{Tomalak:2026btz}     & 0.16 (0.30) & 0.16 (0.30) & 0.16 (0.30) & 0.16 (0.30) & 0.16 (0.30) & 0.16 (0.30) \\ \hline
	total			& 3.1 (0.4) & 3.0 (0.4) & 2.6 (0.4) & 1.7(0.4) & 2.3(0.4) & 2.5 (0.4) \\
	\hline
	\end{tabular}
\end{table}
\begin{figure}[ht]
\begin{center}
	\includegraphics[scale=0.385]{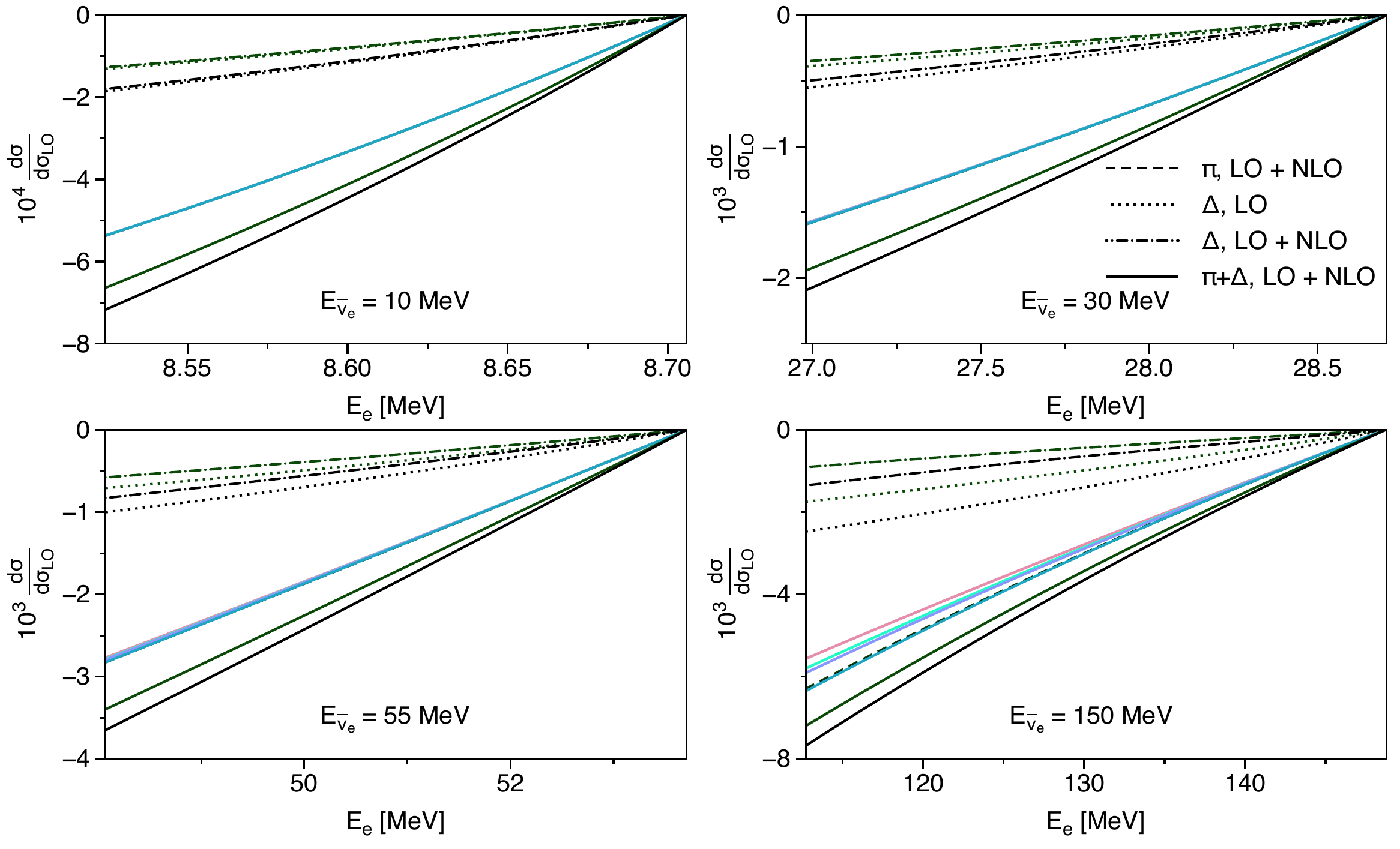}
	\caption{The ratio of the $\Delta$-induced QED radiative contributions to the leading-order cross section $\sigma_{\mathrm{LO}}$ is shown as a function of the recoil positron energy $E_e$ for (anti)neutrino beam energies $E_{\bar{\nu}_e} = 10~\mathrm{MeV}, 30~\mathrm{MeV}, 55~\mathrm{MeV},$ and $150~\mathrm{MeV}$. The dashed lines show the sum of pion-induced QED contributions at LO and NLO (without $\Delta$), the dotted lines show the $\Delta$ contributions at LO, and the sum of LO and NLO $\Delta$ contributions is shown by dash-dotted lines. The total QED radiative corrections in the HBChPT with explicit $\Delta$ degrees of freedom are shown by solid lines. The HBChPT fit parameters are taken from fits to $\pi N$ scattering data~\cite{Siemens:2017opr,Siemens:2020vop}, with constraints from the Roy-Steiner analysis, while the additional constants are the same as in Refs.~\cite{Tomalak:2026wks,Tomalak:2026btz}. The fits are represented by the following colours: $p^4$ in black (the lowest curves), $p^3$ in green (2nd from the bottom). Other fits are illustrated in lighter colours. \label{fig:delta_comparison}}
\end{center}
\end{figure}

We find that including the $\Delta$ resonance in HBChPT improves the convergence of the perturbation theory. $\Delta$ resonance reduce the total QED radiative correction in the ratio $\frac{g_A}{g^\mathrm{QCD}_A g_V}$. Within the uncertainty estimate procedure of Ref.~\cite{Tomalak:2026wks}, i.e., adding in quadrature the uncertainties from the scale variation between $m_n$ and $\sqrt{2} m_n$ of the LO pion-induced QED contributions to nucleon matrix elements and the difference between two evaluation methods for the nucleon contribution to nonperturbative matrix elements, we find that the total pion-induced QED radiative correction is $2.5(0.4)\%$.\footnote{Taking fits in the $K$-matrix approach at $\mathcal{O} \left( p^4 \right)$ instead of the complex mass approach~\cite{Siemens:2020vop}, we obtain $\frac{g_A}{g^\mathrm{QCD}_A g_V} - 1 = 1.9(0.4)\%$ with the LO $\Delta$ contribution $0.3\%$ and the NLO contribution $-0.5\%$.}

\subsection{QED radiative corrections in IBD} \label{subsec:IBD_results}

In figure~\ref{fig:delta_comparison}, we present the ratio of the $\Delta$-induced QED radiative corrections to the leading-order IBD cross section as a function of the recoil positron energy for antineutrino energies $E_{\bar{\nu}_e} = 10~\mathrm{MeV}, 30~\mathrm{MeV}, 55~\mathrm{MeV},$ and $150$ MeV. The kinematic-dependent part of the radiative corrections and the separate contributions start from zero at forward scattering, corresponding to the largest recoil positron energy. As with kinematic-dependent pion-induced contributions, QED radiative corrections from diagrams with virtual $\Delta$ fields are determined by LO HBChPT interactions with negligible effects from the NLO HBChPT Lagrangian up to energies around $150$ MeV. $\Delta$-induced effects are smaller than the contributions from diagrams only with pions. However, these two corrections have the same sign and order of magnitude over all (anti)neutrino energies. Including the $\Delta$ resonance increases the magnitude of the kinematic-dependent QED radiative corrections from diagrams with virtual pions by a factor $1.2$ to $1.3$. The relative contribution of $\Delta$ loops decreases with the (anti)neutrino energy. $\Delta$-resonance contributions are below the uncertainty from the isovector axial-vector form factor~\cite{MINERvA:2023avz,Tomalak:2026wsu} across all energies considered. Precise IBD cross sections at supernova energies and above can be obtained only with improved results for the isovector axial-vector form factor and the HBChPT Wilson coefficients in theories with explicit $\Delta$ degrees of freedom.

\section{Conclusions and Outlook} \label{sec:conclusions}

In this paper, we include contributions from the $\Delta(1232)$ resonance to pion-induced QED radiative corrections in low-energy charged-current processes with nucleons within the framework of heavy-baryon chiral perturbation theory.

Contributions to the vector coupling constant vanish, in agreement with the Behrends-Sirlin-Ademollo-Gatto theorem. Corrections to the isovector axial-vector charge have the same magnitude as corrections from the diagrams with pions only and depend on the leading-order and next-to-leading-order Wilson coefficients $b_4$ and $b_5$. Including the $\Delta$ degrees of freedom improves the HBChPT convergence in the QED radiative correction and reduces QED radiative corrections to the experiment-over-lattice-QCD ratio $\frac{g_A}{g^\mathrm{QCD}_A g_V}$.

Within the precision of our analysis, the $\Delta$-induced QED radiative corrections to the kinematic dependence of IBD cross sections are independent of the NLO HBChPT LECs. $\Delta$ contributions are smaller than corrections with pions only, but have a similar magnitude. $\Delta$ resonance increases the magnitude of the kinematic-dependent QED radiative correction by a factor $1.2$ to $1.3$ compared to the contribution from pions only. The relative correction from $\Delta$ loops decreases with the (anti)neutrino energy and represents the subdominant source of uncertainty in precise predictions of IBD, after errors from the CKM matrix element $V_{ud}$, the isovector axial-vector charge $g_A$, and the isovector axial-vector form factor. Future experimental, phenomenological, and lattice-QCD efforts are essential to reduce these error sources.

\acknowledgments

We thank Zhewen Mo for discussions and his help in preparing the figures while working on the preceding project, and Luis Alvarez Ruso for useful communication. Near completion of this work, a paper~\cite{Gialidi:2026yyk} with partially overlapping research appeared on arXiv, which helped us verify the completeness of relevant interactions and Feynman diagrams. We also acknowledge Emanuele Mereghetti for a subsequent verification of results with~\cite{Gialidi:2026yyk}. This work is supported by the National Science Foundation of China under Grants No. 12347105, 12525504, and 12447101. In this work, FeynCalc~\cite{Mertig:1990an,Hahn:1998yk,Shtabovenko:2016sxi}, Mathematica~\cite{Mathematica}, and DataGraph~\cite{JSSv047s02} were used.

\bibliographystyle{JHEP}
\bibliography{references}

\end{document}